\documentclass[aps,prl,reprint,superscriptaddress,showpacs]{revtex4-1}
\usepackage[utf8]{inputenc}
\usepackage{graphicx}
\usepackage[breaklinks=true,colorlinks=true,linkcolor=blue,urlcolor=blue,citecolor=blue]{hyperref}

\begin{document}


\title{Near-\textit{K}-edge double and triple detachment of the F$^-$ negative ion:\\
observation of direct two-electron ejection by a single photon}

\author{A.~M\"{u}ller}
\email[]{Alfred.Mueller@iamp.physik.uni-giessen.de}
\affiliation{Institut f\"{u}r Atom- und Molek\"{u}lphysik, Justus-Liebig-Universit\"{a}t Gie{\ss}en, 35392 Giessen, Germany}
\author{A.~Borovik Jr.}
\affiliation{I. Physikalisches Institut, Justus-Liebig-Universit\"{a}t Gie{\ss}en, 35392  Giessen, Germany}
\author{S.~Bari}
\affiliation{FS-SCS, DESY, 22607 Hamburg, Germany}
\author{T.~Buhr}
\affiliation{I. Physikalisches Institut, Justus-Liebig-Universit\"{a}t Gie{\ss}en, 35392 Giessen, Germany}
\author{K.~Holste}
\affiliation{I. Physikalisches Institut, Justus-Liebig-Universit\"{a}t Gie{\ss}en, 35392  Giessen, Germany}
\author{M.~Martins}
\affiliation{Institut f\"{u}r Experimentalphysik, Universit\"{a}t Hamburg, 22761 Hamburg, Germany}
\author{ A.~Perry-Sa{\ss}mannshausen}
\affiliation{I. Physikalisches Institut, Justus-Liebig-Universit\"{a}t Gie{\ss}en, 35392  Giessen, Germany}
\author{R.~A.~Phaneuf}
\affiliation{Department of Physics, University of Nevada, Reno, NV 89557, USA}
\author{S.~Reinwardt}
\affiliation{Institut f\"{u}r Experimentalphysik, Universit\"{a}t Hamburg, 22761 Hamburg, Germany}
\author{S.~Ricz}
\affiliation{Institute for Nuclear Research, Hungarian Academy of Sciences, 4001 Debrecen,  Hungary}
\author{K.~Schubert}
\affiliation{FS-SCS, DESY,  22607 Hamburg, Germany}
\author{ S.~Schippers}
\affiliation{I. Physikalisches Institut, Justus-Liebig-Universit\"{a}t Gie{\ss}en, 35392  Giessen, Germany}


\date{\today}

\begin{abstract}
Double and triple detachment of the F$^{-}(1s^2 2s^2 2p^6)$ negative ion by a single photon have been investigated in the photon energy range 660 to 1000~eV. The experimental data provide unambiguous evidence for the dominant role of direct photo-double-detachment with a subsequent single-Auger process in the reaction channel leading to F$^{2+}$ product ions. Absolute cross sections were determined for the direct removal of a $(1s+2p)$ pair of electrons from F$^-$ by the absorption of a single photon.
\end{abstract}




\maketitle



Negative ions are among the most intriguing objects of atomic-scale physics research {~\cite{Andersen2004b}}. Binding an electron to a neutral atom is only possible due to electron correlation, i.e. by electron-electron interactions. The theoretical treatment of such effects has made substantial progress (see e.g.~\cite{Si2017})  since the first applications of quantum mechanics to the description of the H$^-$ ion~\cite{Bethe1929}, but is still difficult, even when only the electronic structure of few-electron atoms or molecules is considered. Treating the effects of electron correlation in processes where negative ions interact with  charged particles or photons provides yet a further challenge to theorists~\cite{Schippers2016a}.

Direct multiple ionization of atoms and atomic ions by a single photon is one of the most fundamental many-body processes. Different from inner-shell excitation with a subsequent cascade of Auger decays, direct photo-double-ionization {(PDI)} is characterized by the absorption of a single photon by an atom and the immediate release of two electrons. This process can solely happen via electron correlation~{\cite{Schneider2002}}.
Thus, {PDI} of atoms is extremely sensitive to the details of the electron-electron interaction and this sensitivity is quadratically enhanced in {PDI} of negative ions.

{There is an increasing interest in direct multiple photoionization for which, throughout this paper, the term photo-multiple-ionization, PMI, (or photo-double-ionization, PDI, in the case of direct ejection of two electrons by one photon) is used. This is in contrast to multiple photoionization which comprises both, direct and sequential processes. For negative ions, ``ionization'' is replaced by ``detachment''. The broad interest in PMI manifests itself in a large body of literature. PMI experiments are conducted predominantly on neutral atoms and molecules~(\cite{Wehlitz2010,Schoeffler2013,Wehlitz2016,Lablanquie2016} and references therein). Although there have been massive theoretical attempts  to calculate total and differential cross sections for PMI of atoms and atomic ions with a focus on the He  and Be  isoelectronic sequences~(\cite{Yerokhin2011,McLaughlin2013,McIntyre2013,Colgan2013,Yerokhin2014,Pindzola2014,Drukarev2016,Bray2017} to name  just a few),  experiments with ions are very scarce. Only few attempts to measure photo-double-detachment, PDD, of negative ions have been reported, and nothing for positive ions. PDD measurements of H$^-$~\cite{Donahue1982}, He$^-$~\cite{Bae1983}, and K$^-$~\cite{Bae1988} were restricted to just a few hundred meV near the PDD threshold. The vigorously pursued theoretical treatments of H$^-$ PDD (\cite{Yip2007} and references therein) could not really be challenged by the H$^-$ experiment~\cite{Donahue1982}.  The theoretical interpretation of a measurement on double photodetachment of F$^-$\cite{Davis2005}  suggested a dominant contribution of PDD at energies around 50~eV, however, the experiment did not yield direct evidence for individual contributions of PDD versus $2s$ inner-valence-shell ionization with subsequent Auger decay.}

There are published experiments on multiple photoionization of positive ions, for example  C$^+$~\cite{Mueller2015a} and {Fe$^+$~\cite{Kjeldsen2002c,Schippers2017}}, and on multiple photodetachment of negative ions, for example {S$^-$~\cite{Bilodeau2005a}, C$^-_{60}$~\cite{Bilodeau2013a} and  O$^-$~\cite{Schippers2016a}. However, in none of these experiments, PMD could be isolated from sequential processes. A previous experiment on Fe$^-$~\cite{Dumitriu2010} did provide evidence for the presence of PDD in the threshold region of (3p+3d) direct two-electron removal but the inferred cross section comprises an additional unknown contribution and could not be followed up to and beyond its maximum.
Obviously, there is a lack of measurements of direct multi-electron removal from an ion by a single photon with a quality that can challenge and guide theoretical treatments. The reason for this situation is that the cross sections are  small and that space charge limits the particle density in a typical ion beam to at most a few times $10^{6}$/cm$^3$ (comparable to ultra-high vacuum) while densities in a gas or vapor target for photoionization experiments can readily reach $10^{13}$/cm$^3$. Reported here is the unambiguous observation of direct double detachment of a $1s$ and a $2p$ electron from the negative ion F$^-$ by a single photon probing inter-shell correlation.  Absolute cross-section measurements for this elusive higher-order process are provided over a significant energy range suitable for guiding and testing theoretical approaches.}

The experimental arrangement and procedures for the present investigation have been previously described in detail~\cite{Schippers2014,Mueller2017}. The photon-ion merged-beams technique was employed using the \textbf{P}hoton-\textbf{I}on spectrometer setup at one of the world's brightest synchrotron-radiation sources, \textbf{PE}TRA III (PIPE).  F$^{-}$ ions were produced in an electron-cyclotron-resonance (ECR) ion source by leaking difluoromethane (CH$_2$F$_2$) gas into the plasma chamber. The ions were accelerated to 6~keV and magnetically analyzed to obtain a pure beam of $^{19}$F$^-$. The ion beam was then transported to the interaction region, collimated and merged with a monoenergetic photon beam  at undulator beamline P04~\cite{Viefhaus2013} of PETRA~III. Product ions were separated from the parent ion beam by a dipole magnet within which the primary beam was collected in a Faraday cup. The photodetached ions were passed through a spherical 180-degree out-of-plane electrostatic deflector to suppress background from stray electrons, photons and ions and then entered a single-particle detector with near-100\% detection efficiency. The photon flux was measured with a calibrated photodiode. {Critical to the measurement of the small cross sections associated with PMD were the high brightness and flux of the photon beam ($3$ to $4\times10^{13}$~s$^{-1}$ at a photon-energy bandwidth of 1 to 1.5~eV in the energy range 660 to 1000~eV), a collimated ion beam with currents as high as 10~nA and low background count rates in the product-ion detector}.

The photon energy scale was calibrated against {Ne (see \cite{Mueller2017}) and O$_2$~\cite{Hitchcock1980a}} reference standards with an estimated uncertainty of  $\pm$0.3~eV. Correction factors of typically 1.0008 were applied to the photon energy in the laboratory frame to account for Doppler shifts due to the counter-propagating ion and photon beams.  The systematic uncertainty of the measured cross sections is estimated to be $\pm$15\% at 90\% confidence level~\cite{Schippers2014}, to which statistical uncertainties in the product ion signal measurements were added in quadrature to give their total uncertainty.

\begin{figure}
\includegraphics[width=\columnwidth]{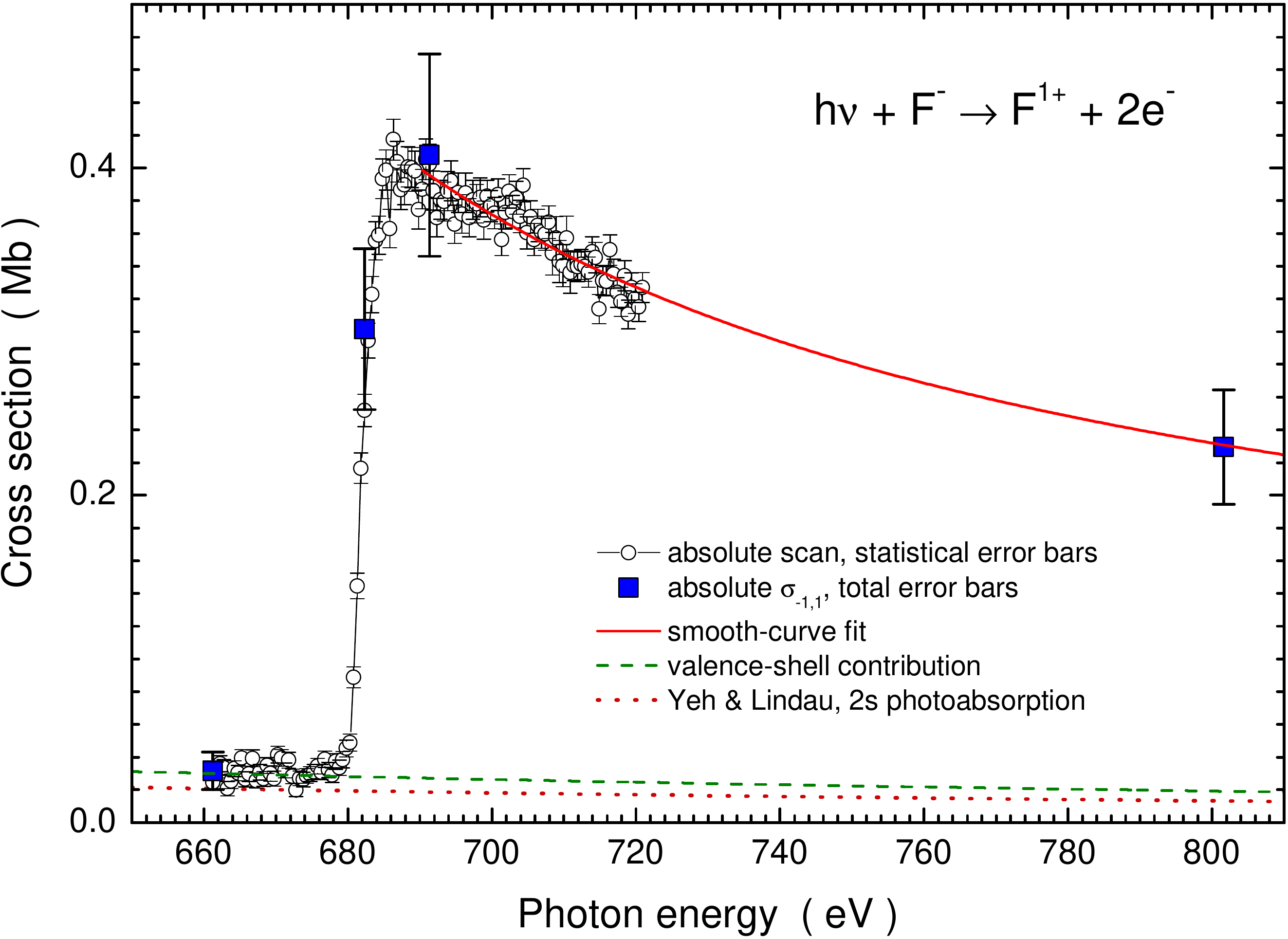}
\caption{\label{Fig:doubledetach} (color online) Absolute cross sections $\sigma_{-1,1}$ for double photodetachment of F$^{-}$ yielding F$^+$ product ions. The open circles with statistical error bars represent a photon-energy scan and the (blue) filled squares indicate absolute measurements with total error bars at fixed photon energies. The solid (red) line is a smooth-curve fit to the measured points. The dotted (dark red) line is the direct-ionization cross section $\sigma_{2s}$ for the $2s$ subshell of neutral fluorine from Yeh and Lindau~\cite{Yeh1985}. The dashed (olive) line  is $1.4 \times \sigma_{2s}$ and is used to simulate the valence-shell contribution to the measured cross section $\sigma_{-1,1}$.}

\end{figure}

 \begin{figure}
\includegraphics[width=\columnwidth]{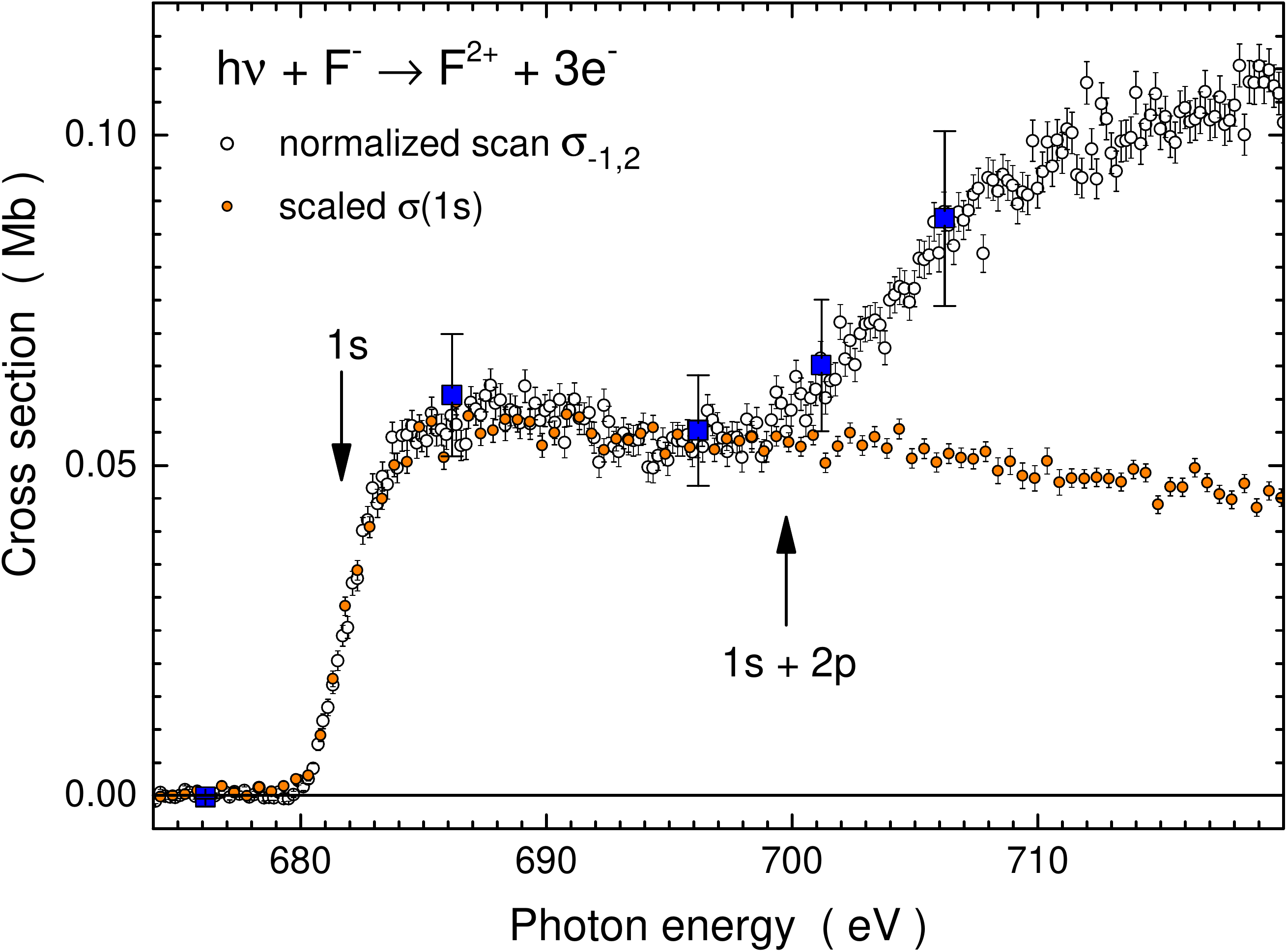}
\caption{\label{Fig:triplescan} (color online) Absolute cross sections $\sigma_{-1,2}$ for triple photodetachment of F$^{-}$ ions. The (blue) filled squares show absolute measurements with total error bars. 
The open circles with statistical error bars were measured in a photon-energy scan and normalized to the absolute measurements.  The smaller circles with (orange) shading were obtained by subtracting the valence-shell contribution from $\sigma_{-1,1}$  (see Fig.~\ref{Fig:doubledetach}) and multiplying the difference by a scaling factor 0.153. They represent the contribution of direct-\textit{K}-shell ionization with subsequent emission of two electrons to $\sigma_{-1,2}$. The arrows indicate the thresholds for $1s$-shell ionization and for the simultaneous removal of a $1s$ and a $2p$ electron from F$^-$.
}
\end{figure}

Figure~\ref{Fig:doubledetach} presents measured cross sections $\sigma_{-1,1}$ for double detachment of F$^-$ ions by a single photon yielding a F$^+$ product ion. At energies above the \textit{K} edge, a dominant contribution from direct \textit{K}-shell photoionization is expected, with a subsequent single-Auger decay, consistent with the measured cross section. The \textit{K}-shell ionization threshold of F$^-$ at about 681~eV is clearly visible and the magnitude of the cross section is close to the threshold value of approximately 0.4~Mb provided by Yeh and Lindau~\cite{Yeh1985} for neutral fluorine. Below the \textit{K} edge, resonances associated with excitation of a $1s$ electron and subsequent decay processes are possible. In the present case, no such resonances were found, indicating that the F($1s2s^22p^6$) core does not support bound states for an additional outer-shell electron. This agrees with findings in the photodetachment of O$^-$ ions where no resonances with principal quantum numbers $n > 2$ were observed~\cite{Schippers2016a}.

Since resonances are absent, double detachment below the \textit{K} edge can only be due to processes involving the valence shell. {PDD releasing two \textit{L}-shell electrons from F$^-$ has been suggested to be dominant at energies around 50~eV~\cite{Davis2005}. At the present energies it is found that  the cross section $\sigma_{2s}$ for $2s$ photoionization of neutral F,  which is shown in Fig.~\ref{Fig:doubledetach} by the dotted (dark red) line, resembles the observed energy dependence of the valence-shell cross-section contribution. Multiplication of this curve by a factor 1.4 gives a good representation of the measured cross section for F$^-$ below the \textit{K} edge which may indicate a dominant contribution of the sequential process at these high energies.} The contribution to the measured cross section above the \textit{K} edge may be estimated by the smooth solid line bridging the gap between the measurements just above the \textit{K} edge and the measured absolute cross section at 800~eV photon energy. This curve is similar in shape and magnitude to the $1s$ photoabsorption cross section of neutral F.

Figure~\ref{Fig:triplescan} shows the result of triple photodetachment measurements with F$^-$ ions at photon energies near the \textit{K} edge. The measured absolute cross-sections $\sigma_{-1,2}$ (filled squares) and normalized scan cross sections (open circles) are compared with the shape of the direct $1s$-ionization contribution (shaded circles) from Fig.~\ref{Fig:doubledetach}. The (orange) shaded circles were obtained by subtracting the valence-shell contribution to $\sigma_{-1,1}$ and multiplying the result by an empirical factor 0.153. They are intended to represent the cross section for direct single \textit{K}-shell ionization with subsequent two-electron emission, assuming a ratio of double- to single-Auger decay of 0.153.
The shaded circles thus obtained give an excellent match with the measured triple-detachment cross section $\sigma_{-1,2}$ up to an energy of approximately 700~eV, above which an additional photoionization mechanism evidently contributes to the cross section. The continuous nature of the additional cross-section contribution and the distinct threshold leave no other explanation than the observation of {PDI} ejecting a pair of inner- and outer-shell electrons.

Threshold energies for the ionization of the $1s$ subshell of F$^-$ obtained from the present experiments and for the removal of a pair of a $1s$ and a $2p$ electron from F$^-$ are indicated by vertical arrows in Fig.~\ref{Fig:triplescan}. The latter threshold is the sum of the binding energy of the extra electron in F$^-(2p^6~^1\textit{S}_0)$, 3.40~eV~\cite{Andersen2004b}, and the \textit{K} edge of neutral F$(2p^5~^2\textit{P}_{3/2})$, {696.8~eV~\cite{Carroll1974}. In the absence of a measurement on atomic F, a  conservative estimate of the uncertainty of this energy is $\pm 2$~eV.   Above the threshold of 700.2~eV the excess cross section rises slowly as expected for PDI in contrast to the  threshold step observed at 681.7~eV in the $1s$ photoabsorption (see Fig.~\ref{Fig:doubledetach}).}

\begin{figure}
\includegraphics[width=\columnwidth]{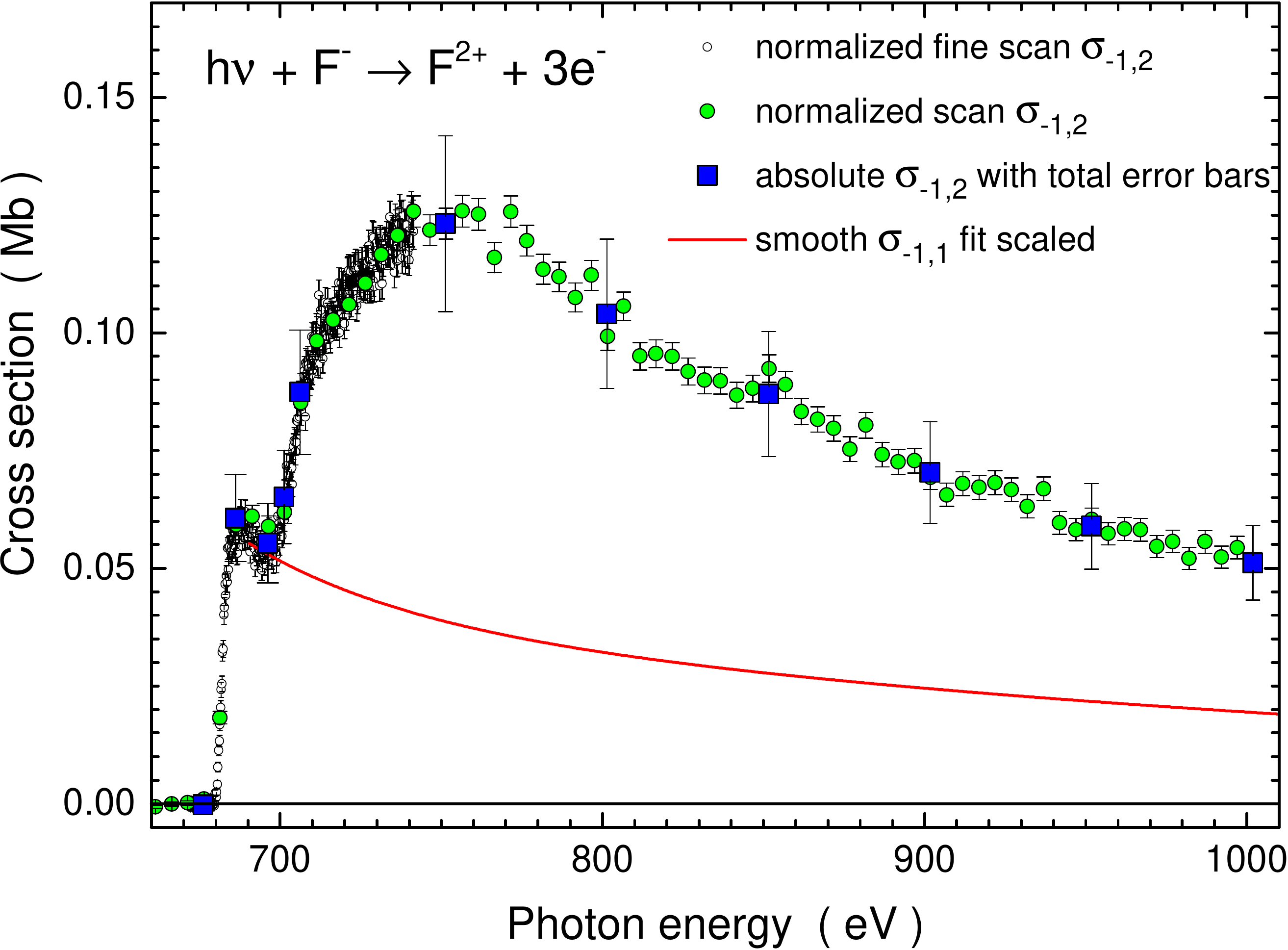}
\caption{\label{Fig:alltriple} (color online) Absolute cross sections $\sigma_{-1,2}$ for triple photodetachment of F$^{-}$ ions over an extended energy range. The open circles with statistical error bars are the normalized scan data from Fig.~\ref{Fig:triplescan}. The (blue) filled squares with total error bars are the results of separate absolute measurements. The (green) shaded circles are normalized scan data taken at 5~eV steps. The solid (red) line is derived from the smooth fit line in Fig.~\ref{Fig:doubledetach} by subtraction of the valence-shell contribution to $\sigma_{-1,1}$  (see Fig.~\ref{Fig:doubledetach}) and multiplying the difference by an empirical factor of 0.153.
}
\end{figure}

In order to further investigate the additional process contributing to $\sigma_{-1,2}$, the photon energy range of the measurements was extended to 1000~eV. Figure~\ref{Fig:alltriple} shows the results of an additional photon-energy scan covering the energy range 660 to 1000~eV in steps of 5~eV together with absolute measurements to which the scan was normalized. The results for $\sigma_{-1,2}$ from Fig.~\ref{Fig:triplescan} are included for completeness. Also shown is a solid (red) line which is a smooth representation of the cross section contribution of direct $1s$ ionization with subsequent double-Auger decay. This line was obtained by subtracting the valence-shell contribution to $\sigma_{-1,1}$ from the solid smooth line in Fig.~\ref{Fig:doubledetach} that essentially represents $\sigma_{-1,1}$ at photon energies beyond 690~eV and multiplying the result by a factor 0.153. Thus a smooth extrapolation of the shaded circles shown in Fig.~\ref{Fig:triplescan} is available for the further analysis of the measured data.

\begin{figure}
\includegraphics[width=\columnwidth]{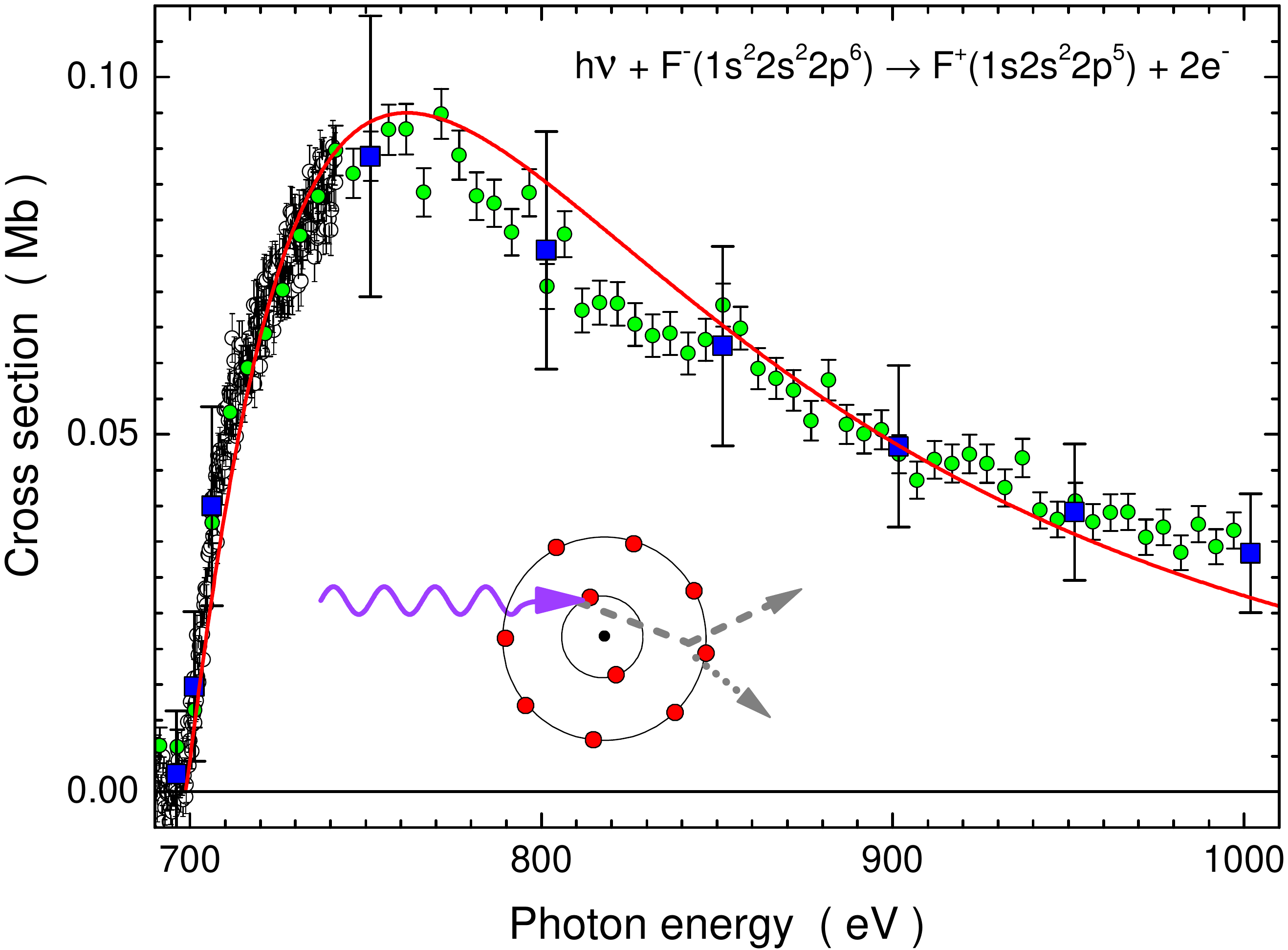}
\caption{\label{Fig:directdouble} (color online) The difference (divided by 0.95; see text) of the measured data points and the solid line shown in Fig.~\ref{Fig:alltriple}. This difference represents the absolute cross section for direct double-detachment of F$^-$ ions, simultaneously removing a pair of a $1s$ and a $2p$ electron with a subsequent Auger decay  releasing a third electron. The solid (red) line models  the PDD cross section on the basis of the Pattard scaling rule~\cite{Pattard2002}. The inset shows a cartoon of {PDD} of the initial F$^-$ ion.}
\end{figure}

It is now possible to extract the partial contribution to the total triple photodetachment cross section that arises from {PDD} of the F$^-$ ion releasing  a $1s$ and  $2p$ electron pair
\begin{equation}
h\nu + \textrm{F}^-(1s^2 2s^2 2p^6) \to \textrm{F}^+(1s 2s^2 2p^5) + 2e^-
\end{equation}
 and a subsequent single-Auger decay
 \begin{equation}
\textrm{F}^+(1s 2s^2 2p^5) \to \textrm{F}^{2+}(1s^2 2s^2 2p^3) + e^- .
\end{equation}
The probability for single-Auger decay of F$^+(1s 2s^2 2p^5)$ is approximately $0.95 \pm 0.05$, as demonstrated in separate experiments with F$^+$ ions~\cite{Reinwardt2018}. Hence, the difference between the measured triple-detachment cross sections $\sigma_{-1,2}$ and the estimated contribution arising from $1s$ photoionization with subsequent double-Auger decay (the solid line in Fig.~\ref{Fig:alltriple}) accounts for nearly the entire cross section for {PDD} of a pair of a $1s$ and a $2p$ electron from the F$^-$ ion. This cross section, corrected for a single-Auger branching ratio of 0.95, is shown in Fig.~\ref{Fig:directdouble}. The magnitude of the {PDD} cross section is 95~kb at its maximum. Its shape as a function of photon energy is that expected for a {PMI} process. This can be investigated on the basis of the {Wannier threshold law~\cite{Wannier1953} and the} scaling rule constructed by Pattard~\cite{Pattard2002}. {The threshold region of the cross section $\sigma(E)$ in Fig.~\ref{Fig:directdouble} was investigated using the Wannier threshold law $\sigma(E) = \sigma_0 (E - E_{th})^\alpha$. For energies up to 707.6~eV $\sigma(E)$ could be fitted with a constant $\sigma_0$, $E_{th} = 698.7(1.5)$~eV, and $\alpha = 1.1(3)$. This is compatible with the expected Wannier exponent for PDD from F$^-$, $\alpha = 1.1269$, and the threshold energy $E_{th} = 700.2$~eV.}

The scaling of cross sections derived by Pattard for {PDI and PDD}  is expressed by
\begin{equation}
\sigma(E) = \sigma_M \, x^\alpha \left( \frac{\alpha+7/2}{\alpha x + 7/2}  \right) ^{(\alpha+7/2)}
\end{equation}
with the photon energy $E$, the maximum cross section $\sigma_M$,  $x=(E-E_{th})/(E_M-E_{th})$, $E_M$ the energy where the cross section reaches its maximum,
{and the Wannier exponent $\alpha$~\cite{Wannier1953}.
The solid (red) line in Fig.~\ref{Fig:directdouble} shows the Pattard scaling for the fitted parameters $\sigma_M = 0.095$~Mb, and $E_M = 763.1$~eV with $\alpha = 1.1$ and $E_{th} = 698.7$~eV fixed.
Given the uncertainty of the cross-section measurement and that of the subtracted partial cross section (the solid line in Fig.~\ref{Fig:alltriple}) the scaling suggested by Pattard agrees remarkably well with the experimentally derived cross section in Fig.~\ref{Fig:directdouble}, supporting the conclusion that indeed PDD of F$^-$ has been observed.}

While one would expect direct \textit{K}-shell single photoionization with subsequent single- or double-Auger decays to be the dominant mechanisms for the removal of two or three electrons from an atom or ion, the present experiment shows that {PDD} of F$^-$ is responsible for the dominant contribution to the cross section $\sigma_{-1,2}$ for triple detachment by a single photon. A pair of electrons, one from the $1s$ shell and one from the $2p$ subshell, are released  and subsequently, with a probability of approximately 95\%, a single-Auger decay produces the final charge state F$^{2+}$. The dominance of the {PDD} contribution to the total triple-detachment cross section is attributed to the small binding energy of the outermost $2p$ electron in  the F$^-$ parent ion. The capability of the present experiment to differentiate between the charge states of the photoions produced after absorption of a  photon facilitated the clear observation of a process that is characterized by a very small cross section, but dominantly contributes to the production of F$^{2+}$ ions via net triple detachment of F$^-$.

 The present measurement has opened a window to an additional dimension in experimental studies of direct multiple ionization by a single photon. This dimension is the charge state $q$ of atoms/ions which can be varied, for example, along isoelectronic sequences with a fixed number of electrons $Z-q$ where $Z$ is the atomic number of the investigated element. While previous experiments were restricted to neutral species such as the helium atom, the extension to measurements with ions allows one to manipulate the relative strengths of the electron-electron and electron-nucleus interactions, thereby facilitating a systematic variation of the essential forces governing the  structure of atoms and ions and their dynamical response  to external perturbations. {Preliminary experiments with Ar$^+$ ions at the present setup indicate the feasibility of cross-section measurements for PDI of \textit{L}-shell electrons in a positive ion. Results similar to the present one are expected with other negative ions.} Measurements of cross sections for PDI of isoelectronic atomic ion species will help to better understand the balance between electron correlation and electron-nucleus interactions in atomic structure and interactions.

This research was carried out in part at the light source PETRA III at DESY, a member of the Helmholtz Association (HGF).
Support from Bundesministerium f\"{u}r Bildung und Forschung provided within the "Verbundforschung" funding scheme (contract numbers 05K10RG1,  05K10GUB, 05K16RG1, 05K16GUC) and from Deutsche Forschungsgemeinschaft under project numbers Mu 1068/22, Schi 378/12, and SFB925/A3  is gratefully acknowledged. R.P. acknowledges support from the Alexander von Humboldt Foundation. S.B. and K.S. were supported by the Helmholtz Initiative and Networking Fund through the Young Investigators Program and by the Deutsche Forschungsgemeinschaft, project B03/SFB755. We  thank K. Bagschik, J. Buck, F. Scholz, J. Seltmann, and J. Viefhaus for assistance in using beamline P04.


%

\end{document}